\documentclass[aps,prl,showpacs,twocolumn,superscriptaddress]{revtex4-1}
\usepackage{amsmath}
\usepackage{amssymb}
\usepackage{graphicx}
\usepackage{hyperref}
\usepackage{url}
\usepackage{color}
\usepackage{ulem}
\begin{document}
\title{BaFe$_2$Se$_3$: a high $T_{\rm C}$ magnetic multiferroic with large ferrielectric polarization}
\author{Shuai Dong}
\affiliation{Department of Physics, Southeast University, Nanjing 211189, China}
\author{J.-M. Liu}
\affiliation{National Laboratory of Solid State Microstructure, Nanjing University, Nanjing 210093, China}
\author{Elbio Dagotto}
\affiliation{Department of Physics and Astronomy, University of Tennessee, Knoxville, Tennessee 37996, USA}
\affiliation{Materials Science and Technology Division, Oak Ridge National Laboratory, Oak Ridge, Tennessee 37831, USA}
\date{\today}

\begin{abstract}
The iron-selenides are important because
of their superconducting properties. Here, an unexpected
phenomenon is predicted to occur in an iron-selenide compound with a
quasi-one-dimensional ladder geometry: BaFe$_2$Se$_3$ should be
a magnetic ferrielectric system, driven by its magnetic block order via
exchange striction. A robust performance
(high $T_{\rm C}$ and large polarization) is expected.
Different from most multiferroics,
BaFe$_2$Se$_3$ is ferrielectric, with a polarization that
mostly cancels between ladders. However, its strong magnetostriction still produces
a net polarization that is large ($\sim$$0.1$ $\mu$C/cm$^2$) as compared
with most magnetic multiferroics. Its fully ferroelectric state, with
energy only slightly higher than the ferrielectric, has
a giant improper polarization $\sim$$2-3$ $\mu$C/cm$^2$.
\end{abstract}
\maketitle

\textit{Introduction}.
Low critical temperatures ($T_{\rm C}$'s) and weak ferroelectric (FE)
polarizations ($P$'s) are two important drawbacks of current
type-II multiferroics, where $P$'s are
driven by magnetism~\cite{Khomskii:Phy,*Cheong:Nm,*Wang:Ap,*Dong:Mplb}.
For this reason, a considerable effort recently
focused on the design of new magnetic multiferroics
to improve on $T_{\rm C}$ and its associated FE $P$. A recently
confirmed example involves the quadruple-perovskite manganite CaMn$_7$O$_{12}$, with
relatively large $P$ ($\sim0.3$ $\mu$C/cm$^2$) and $T_{\rm C}$
($90$ K) \cite{Zhang:Prb11,*Johnson:Prl}, triggered by a new multiferroic
mechanism \cite{Lu:Prl,Dong:Prl,*Perks:Nc,*Mostovoy:Phy}.

Despite the conceptual differences between superconductivity and multiferroicity,
the search for high $T_{\rm C}$ superconductors (SCs) can help
the magnetoelectric (ME) community to develop multiferroics with even higher
$T_{\rm C}$'s. For example, Kimura \textit{et al.} found that CuO
(a material related to Cu-oxide SCs)
is actually a high-$T_{\rm C}$ type-II multiferroic between $213$-$230$ K \cite{Kimura:Nm}. Besides the cuprates, the iron-based pnictides and chalcogenides have been intensively studied since 2008 because of their
superconducting properties~\cite{Johnston:Ap,*Stewart:Rmp,*Dagotto:Rmp}.
However, to our knowledge the possibility of multiferroic
behavior has not been investigated before in any of these systems.

In this Letter, the iron-selenide
BaFe$_2$Se$_3$ is predicted to hide a robust
multiferroic order.
Until now, BaFe$_2$Se$_3$ has been investigated
as a member of the Fe-based superconductors
family with only a handful of efforts that focused
on magnetism~\cite{Maziopa:Jpcm,*Svitlyk:Jpcm,Caron:Prb,*Caron:Prb12,Nambu:Prb,Saparov:Prb,Medvedev:Jetp,Lei:Prb}
and (unconfirmed) superconductivity.
Our prediction instead provides a novel and unexpected perspective of BaFe$_2$Se$_3$, that potentially
may extend the search for multiferroics beyond this compound into
the chalcogenides/pnictides families with tetrahedral
anion cages.

\begin{figure*}
\centering
\includegraphics[width=0.9\textwidth]{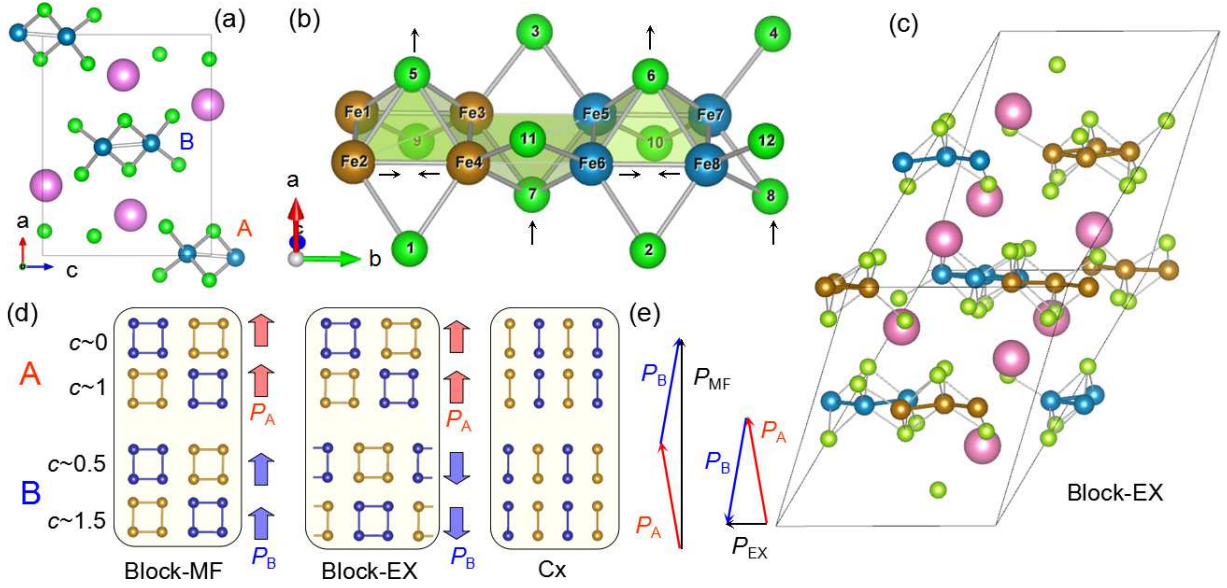}
\caption{(Color online) Crystal and magnetic structures of BaFe$_2$Se$_3$. (a)
Side view along the $b$-axis. Blue: Fe; green: Se; pink: Ba. (b) A Fe-Se
ladder along the $b$-axis and its magnetic order.
Partial ionic displacements driven by the exchange striction are marked as black arrows.
(c) A unit cell considering
the AFM magnetic order. (d) Spin structures. Left: Block-MF; middle: Block-EX; right: Cx. The side
arrows denote the local FE $P$'s of each ladder. In (b-d), the spins ($\uparrow$/$\downarrow$) of
Fe's are distinguished by colors. (e) Vector addition of FE $P$'s of ladders A
and B.
}
\label{struc}
\end{figure*}

BaFe$_2$Se$_3$ forms an orthorhombic structure. Each unit cell has two iron ladders (labeled as
A and B), built by edge-sharing FeSe$_4$ tetrahedra, as
shown in Fig.~\ref{struc}(a-b). Long-range antiferromagnetic (AFM) order is
established below $256$ K \cite{Caron:Prb}. Both neutron studies and
first-principles calculations reported
an exotic block AFM order~\cite{Caron:Prb,Saparov:Prb,Nambu:Prb,Medvedev:Jetp} [Fig.~\ref{struc}(b-c)]. The Hartree-Fock approximation to the five-orbital
Hubbard model also confirmed the stability of the block AFM phase and revealed
other competing phases, e.g.
the Cx phase [Fig.~\ref{struc}(d)]~\cite{Luo:Prb}.

\textit{Symmetry analysis}.
The block AFM order is particularly interesting because it breaks parity symmetry and displays exchange striction effects. Indeed, the iron displacements are
prominent, as revealed by neutron
studies~\cite{Maziopa:Jpcm,Caron:Prb,Saparov:Prb,Nambu:Prb,Caron:Prb12}:
the nearest-neighbor (NN) distances between
Fe($\uparrow$)-Fe($\uparrow$) [or Fe($\downarrow$)-Fe($\downarrow$)] at $200$ K
become $2.593$ \AA{}, much shorter than the Fe($\uparrow$)-Fe($\downarrow$) distance
$2.840$ \AA{} \cite{Caron:Prb12}. However, this exchange striction is not sufficient
to induce FE $P$ since it breaks
parity but not space-inversion symmetry. Thus,
although neutron studies reported
exchange striction effects in iron ladders \cite{Maziopa:Jpcm,Caron:Prb,Saparov:Prb,Nambu:Prb,Caron:Prb12},
ferroelectricity has not been searched for in these materials.

The Se-tetrahedra also
break parity in each ladder since Fig.~\ref{struc}(b) shows that
Se(5) is above the ladder's plane but the next Se(7) is below,
and the distances of Se(5) and Se(7) to the iron ladder plane should
be the same in magnitude and opposite sign (``antisymmetric'').
However, the 
block AFM order 
introduces
a fundamental modification in the symmetry. 
Now the blocks made of four Fe($\uparrow$)'s [or four Fe($\downarrow$)'s] are no
longer identical to blocks made of two Fe($\uparrow$)'s and
two Fe($\downarrow$)'s. Then, the Se(5) and
Se(7) heights do not need to be antisymmetric anymore; their distances
to the ladder planes can become different.
The same mechanism works for the edge Se's, e.g.
Se(1) and Se(11). As a consequence, the Se atomic positions
{\it break} the space inversion symmetry, generating a local FE
$P$ pointing perpendicular to the iron ladders plane (almost along the $a$-axis).
Previous neutron studies~\cite{Caron:Prb} could have observed this effect,
but in those investigations the Se positions were not discussed since the
focus was not multiferroicity. 
Similar exchange striction works in the E-type AFM
manganites 
and in Ca$_3$CoMnO$_6$ 
although the details
are not identical \cite{Sergienko:Prl,Choi:Prl}.

Qualitatively, the ME coupling energy for each ladder \cite{Sergienko:Prl} can be analytically expressed as:
\begin{equation}
F=\alpha(\textbf{B}_1^2-\textbf{B}_2^2)P_\perp+\frac{1}{2\chi}\textbf{P}^2,
\label{energy}
\end{equation}
with the parity order parameters $\textbf{B}_1=\textbf{S}_1+\textbf{S}_2+\textbf{S}_3+\textbf{S}_4-\textbf{S}_5-\textbf{S}_6-\textbf{S}_7-\textbf{S}_8$;  $\textbf{B}_2=\textbf{S}_1+\textbf{S}_2-\textbf{S}_3-\textbf{S}_4-\textbf{S}_5-\textbf{S}_6+\textbf{S}_7+\textbf{S}_8$. $\textbf{S}_i$ denotes the spin of Fe($i$) as indicated in Fig.~\ref{struc}(b). $\alpha$ is the coefficient of exchange striction,
proportional to $\frac{\partial J}{\partial r}$ where $J$ and $r$ are the exchange and distance between NN Fe's along the ladder direction, respectively.
$\chi$ is the dielectric susceptibility of the paraelectric phase.
$P_\perp$ is the FE component perpendicular to the Fe ladder plane.
By minimizing the energy, the induced $P$ of each ladder can be
obtained as $-\alpha\chi(\textbf{B}_1^2-\textbf{B}_2^2)$, perpendicular
to the ladder plane. This scheme is similar (but not identical) to
that of E-type AFM $o$-HoMnO$_3$ \cite{Sergienko:Prl}, and different
in principle from geometric improper
ferroelectrics \cite{Bousquet:Nat,*Benedek:Prl,*Yang:Prl}.

This discussion suggests that each ladder can be multiferroic, but only the inclusion of
inter-chain couplings can address if a macroscopic FE $P$ will indeed be generated.
According to neutron studies \cite{Caron:Prb},  the block AFM pattern shows a
$\frac{\pi}{2}$-phase shift between the NN A-B ladders but a $\pi$-phase
shift between the NN A-A ladders (and NN B-B ladders), as in the Block-EX shown in
Fig.~\ref{struc}(d). Then, the unit cell of BaFe$_2$Se$_3$ doubles when
considering the magnetism [see Fig.~\ref{struc}(c)].
According to the analytical expression above,
the $\pi$-shift between A-A ladders (or B-B ladders)
will not change the direction of the induced FE $P$ \footnote{Note that all the ladders, either A or B, have the same staggered pattern of Se atoms above and below the ladder planes, i.e. a plaquette with
a Se(5) above the ladder plane for ladder A, corresponds to another Se(5) above the ladder plane
for ladder B, for the same plaquette.} 
but the $\frac{\pi}{2}$-phase shift between A-B ladders will induce (nearly)
opposite FE $P$'s, as sketched in Fig.~\ref{struc}(d-e).
A full cancellation does not occur due to a second key observation:
a small canting angle exists between the
ladders A and B planes [see Fig.~\ref{struc}(a)],
leading to a residual FE $P$ ($P_{\rm EX}$)
pointing almost along the $c$-axis [Fig.~\ref{struc}(e)].
The residual $P_{\rm EX}$ magnitude can be
estimated by considering the tilting
angle between the ladders A and B planes, which is about $5.4^\circ$
according to experiments~\cite{Caron:Prb}. This small
tilting gives $P_{\rm EX}\approx9.4\%P_A$.

Since the spin ladders in BaFe$_2$Se$_3$ are
quasi-one-dimensional, the inter-ladder couplings should be weak compared
to the intra-ladder couplings. Thus, it may be possible to overcome the
$\frac{\pi}{2}$-phase shift between ladders A and B by
chemical substitution, or electric field.
If this is achieved, the magnetic structure becomes the Block-MF state. In this case, the magnetism-induced FE $P$'s of all ladders will
coherently produce a combined $P_{\rm MF}$ pointing along the $a$-axis
[Fig.~\ref{struc}(e)], with an amplitude nearly twice that of $P_A$.
All this intuitive analysis for the many possible magnetic states has
been fully confirmed by formal group theory
\footnote{
Using the experimental structure (space group No. 62, $Pnma$, orthorhombic)
plus the particular magnetic order, the magnetic space group becomes monoclinic:
(1) No. 14, $P21/c$ for the Cx phase; (2) No. 9, $Cc$ for the Block-EX phase;
(3) No. 8, $Cm$ for the Block-MF phase. The point group of $P21/c$ is $2/m$
which is a nonpolar point group. Then, $P$ is forbidden in this
group. The point group of both $Cc$ and $Cm$ is $m$, which is a polar point
group and allows $P$.}.

\begin{figure}
\centering
\includegraphics[width=0.48\textwidth]{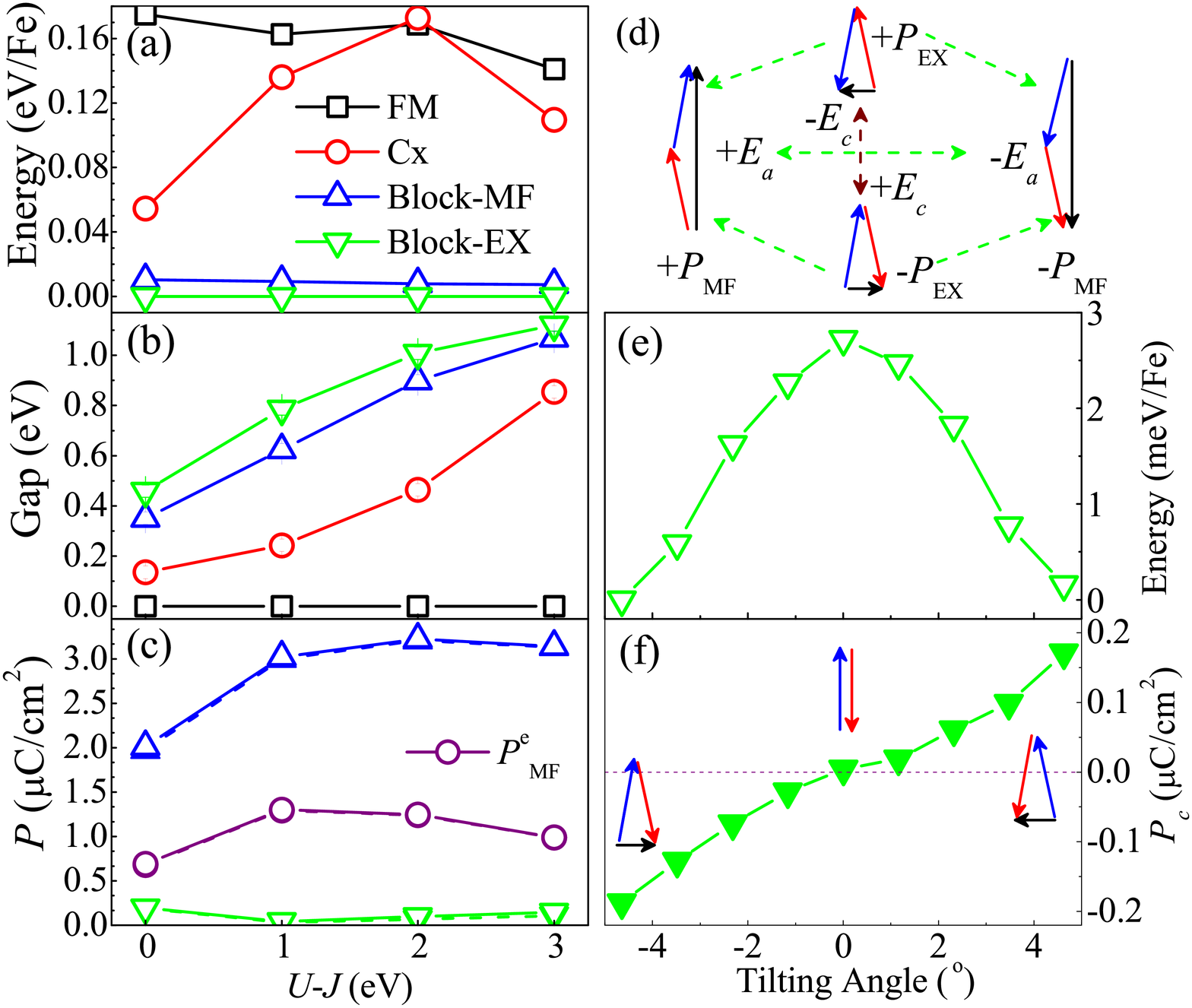}
\caption{(Color online) (a-c) DFT results varying the effective Hubbard
interaction. (a) Energies for various magnetic states, with the Block-EX
as reference. 
(b) Band gaps. NM and FM are metallic (zero gap).
(c) The FE $P$'s of Block-MF and Block-EX states. The dashed lines (with solid symbols) are the components
 along the symmetry expected directions (e.g. $a$ axis for Block-MF, $c$ axis for Block-EX),
which are almost identical to the total $P$ and imply a successful prediction by the symmetry analysis. The purple $P_{\rm MF}^{\rm e}$'s (solid and open symbols nearly overlapping) are the pure electronic contribution in the Block-MF case.
(d) Sketch of switchings between $\pm P_{\rm EX}$ and $\pm P_{\rm MF}$ driven by the electric field $E_x$ along the $x$ ($x$=$a$ or $c$) direction.
(e-f) DFT demonstration (without $U$) of switching between $\pm P_{\rm EX}$ via the rotation of Fe-ladder planes. Horizontal axis: the angle between ladders A's and B's planes. The two limits ($\sim\pm4.6^\circ$) denote the relaxed $\pm P_{\rm EX}$ states, respectively. The center $0^\circ$ denotes the relaxed non-tilting case. For other angles, the structures are obtained by proportional mixing among these three limits. Vertical axes: (e) energy per Fe; (f) polarization along the $c$-axis.}
\label{dft}
\end{figure}

\textit{First-principles study}.
A density functional theory (DFT) calculation will
be used to confirm above predictions \cite{Supp}.
The DFT results are in Fig.~\ref{dft} varying
the effective Hubbard interaction $U-J$, which give the following conclusions:

{\it (1)} Atomic positions were optimized with the relevant
magnetic states [ferromagnetic (FM), Cx-type AFM, Block-MF, Block-EX,
and non-magnetic (NM)],
and their energies were compared. As shown in Fig.~\ref{dft}(a), the
Block-EX state is the lowest in energy, as in
experiments~\cite{Maziopa:Jpcm,Caron:Prb,Saparov:Prb,Nambu:Prb,Caron:Prb12}.
The Block-MF 
is only slightly higher
($7$-$10$ meV/Fe). All other states are much higher.
In the Block-EX state, the DFT NN Fe-Fe distance for
Fe($\uparrow$)-Fe($\uparrow$) [or Fe($\downarrow$)-Fe($\downarrow$)]
is $\sim 2.58$$4$ \AA{} and for Fe($\uparrow$)-Fe($\downarrow$)
$\sim 2.823$ \AA{} (without $U$), very similar to the neutron results mentioned before.
More importantly, DFT finds that the heights of Se(5) and Se(7) are
different: 1.64 \AA{} and 1.42 \AA{} (without $U$), respectively. This numerically confirms
that the relaxed structures of the individual ladders do have
a net electric moment.

{\it (2)} The density of states (DOS) were calculated to extract the energy
gap around the Fermi level. As shown in
Fig.~\ref{dft}(b), the FM and NM states are metallic while all other
magnetic states are insulating. 
Note that the Block-EX energy gap is $0.46$ eV without $U$, in agreement
with previous DFT results ($0.44$ eV \cite{Saparov:Prb}) but much higher than the
value estimated from the resistance-temperature curves which is
$\sim 0.13$-$0.178$ eV \cite{Nambu:Prb,Lei:Prb}. This difference is
important and will be further discussed below.


{\it (3)} The most important physical property is the FE $P$. The insulating
and space-inversion symmetric Cx phase is considered as the nonpolar reference
state. Confirming the previous symmetry analysis,
both the Block-EX and Block-MF are found to be multiferroic
in our DFT calculations. $P_{\rm MF}$ is large
and mostly along the $a$-axis ($2.01$ $\mu$C/cm$^2$ without $U$ and increases to $3.02-3.22$ $\mu$C/cm$^2$ with $U$). This value of $P_{\rm MF}$ is among the largest reported in type-II
multiferroics, comparable with the E-type AFM
manganites~\cite{Sergienko:Prl,Picozzi:Prl,Nakamura:Apl}. As discussed
before, the Block-EX should be ferrielectric with a weaker $P$. This is also confirmed
in our DFT calculation: the net FE $P$ is mostly along the $c$-axis and its
amplitude is $0.19$ $\mu$C/cm$^2$ without $U$, which is one order of magnitude smaller than $P_{\rm MF}$ as expected from the above symmetry analysis, and comparable with $R$MnO$_3$ ($R$= Tb or Dy) \cite{Kimura:Prb05} The $U$-dependent $P_{\rm EX}$ is non-monotonic \cite{Supp}.
The DFT directions of $P_{\rm EX}$ and $P_{\rm MF}$
agree perfectly with the symmetry analysis, and the values of $P_{\rm EX}$ and $P_{\rm MF}$
are also in qualitative agreement.

{\it (4)} Although the experimental-measurable quantity is the total $P$,
it is physically meaningful to analyze the individual contributions from
ionic and electronic displacements. Previous DFT studies on type-II
multiferroics reported that the electronic contribution could be significant~\cite{Picozzi:Prl,Lu:Prl},
contrary to proper ferroelectrics where
the ionic displacements are always dominant. Thus, it is
interesting to disentangle the electronic $P^{\rm e}$ and ionic $P^{\rm ion}$
contributions in BaFe$_2$Se$_3$. To unveil
the intrinsic physics of each ladder and avoid compensation effects between ladders,
here the Block-MF case is analyzed. By adopting the relaxed structure with the
Cx magnetic order and imposing the Block-MF spin order, the pure electronic
contribution $P_{\rm MF}^{\rm e}$ can be estimated: it results to be
large ($\sim0.69$-$1.3$ $\mu$C/cm$^2$, about $1/3$ of $P_{\rm MF}$ and parallel to $P_{\rm MF}$.

{\it (5)} As sketched in Fig.~\ref{struc}(d), by shifting the magnetic blocks by one lattice constant
in all ladders, both $P_{\rm A}$ and $P_{\rm B}$ are reversed according
to the analytical formula above. Then both $P_{\rm EX}$ and $P_{\rm MF}$
can be flipped by $180^\circ$. The energies before and after such a $180^\circ$ flipping
are degenerate. As sketched in Fig.~\ref{dft}(d), to realize the flipping of $P_{\rm EX}$,
an external electric field should be applied along the $c$-axis. If a large enough field
is applied along the $a$-axis, the ferrielectric (Block-EX) to FE (Block-MF) phase transition will occur, producing a $90^\circ$ flipping and enhancement of $P$. Moreover,
the $180^\circ$ flipping of $P_{\rm EX}$ can also been obtained by reversing
the titling angle between the planes of ladders A-B, without shifting the magnetic blocks. As shown in Fig.~\ref{dft}(e), The calculated energy shows an almost
symmetric barrier between the $+P_{\rm EX}$ and $-P_{\rm EX}$ states, 
with the height of the barrier of $\sim2.8$ meV/Fe. Comparing with other FE materials, e.g. $8$ meV/Mn for $o$-HoMnO$_3$ and $18$ meV/Ti for BaTiO$_3$ \cite{Picozzi:Prl}, the required electric fields $\pm E_c$ should be accessible. Note that this switching path is an energetically ``upper bound'', not necessarily the actual path occurring in experiments during switching, which may display an even lower energy barrier.
In addition, a magnetic field can suppress the AFM order and its FE $P$, as in other spin-$\uparrow\uparrow\downarrow\downarrow$ multiferroics (e.g. Ca$_3$CoMnO$_6$ \cite{Choi:Prl}), rendering an intrinsic ME coupling.

In summary, our DFT calculations fully confirm the proposed
magnetic-induced ferrielectricity of BaFe$_2$Se$_3$.
The multiferroic properties of BaFe$_2$Se$_3$ are very
prominent: (1) high $T_{\rm C}$ close to room temperature;
(2) large polarization in the ground state and even larger
in the excitation state. Both these two properties are in
the topmost range among all type-II multiferroics, i.e. BaFe$_2$Se$_3$
can be a quite interesting material.

\textit{Additional discussion}.
Since pure DFT always
underestimates the band gap, the real band gap of
BaFe$_2$Se$_3$ should be even larger, and the observed small
gaps ($0.13$-$0.178$ eV \cite{Nambu:Prb,Lei:Prb}) in transport may be caused by in-gap levels induced
by impurities. In fact, non-stoichiometry and impurities are ubiquitous in all samples of
BaFe$_2$Se$_3$ in previous experiments~\cite{Caron:Prb,Saparov:Prb,Lei:Prb,Caron:Prb12,Nambu:Prb},
making these samples too conductive to detect ferro- or ferri-electricity.

To guide future experimental efforts, here results for the
iron-selenides BaFe$_2$S$_3$ and KFe$_2$Se$_3$ are also presented.
Although BaFe$_2$S$_3$ is very similar to
BaFe$_2$Se$_3$, its space group is the orthorhombic $Cmcm$~\cite{Hong:Jssc},
identical to that of KFe$_2$Se$_3$ \cite{Caron:Prb12}.
Furthermore, our DFT calculation on BaFe$_2$S$_3$ predicts
a Cx ground state as found in KFe$_2$Se$_3$, in agreement with recent experiments \footnote{H. Takahashi, private communication.}.
Considering the magnetic similarity
between BaFe$_2$S$_3$ and KFe$_2$Se$_3$, it is reasonable to assume that the
Fe-Se bond in the latter may not be fully electrovalent due to the weak
electronegativity of Se. 
In this sense, the real
Fe valence in BaFe$_2$Se$_3$ is $1+\delta$ (with $\delta$ between 0 and 1)
instead of the nominal $+2$, which
may be the reason for the experimental difficult to prepare pure BaFe$_2$Se$_3$ due
to the instability of Fe$^{(1+\delta)+}$, which will induce iron vacancies \cite{Saparov:Prb}.
Even the exotic AFM block state, with
tetramer magnetic units, may be also caused by this $1+\delta$ effect
according to the mechanism of
Peierls-like transition in one-dimensional lattices, e.g. at $\delta$=$\frac{3}{4}$
or $\delta$=$\frac{1}{4}$.

\begin{figure}
\centering
\includegraphics[width=0.45\textwidth]{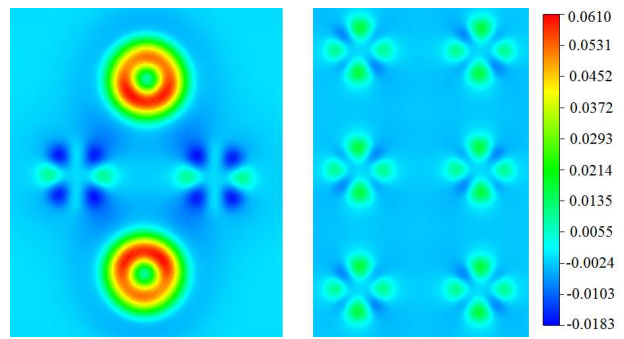}
\caption{(Color online) Two-dimensional profiles of electronic density difference (BaFe$_2$S$_3$ minus BaFe$_2$Se$_3$). Left: the Se(5)-Fe(3)-Fe(4)-Se(7) plane. Spheres denote the Se/S sites while multi-lobe ones the Fe sites. Right: the Fe-ladder plane.}
\label{charge}
\end{figure}

The argument above is clear in our DFT calculation. The electron density differences
between BaFe$_2$Se$_3$ and BaFe$_2$S$_3$ are displayed in Fig.~\ref{charge}.
The bright red spheres provide clear evidence that the S anions attract more
electrons than Se. Meanwhile, the Fe cations lose more $3d$ electrons
in BaFe$_2$S$_3$, characterized by bright blue lobes pointing along
the Fe-S/Se directions. By contrast, the density difference is weak
but also exists in the Fe-Fe ladder plane.
Besides these two clear differences, outside the bright
green spheres, there is a dim blue sphere surrounding each S/Se site,
with negative value: this suggests that the outmost
electrons of Se (S) are more extended (localized), also supporting
the covalent scenario for BaFe$_2$Se$_3$.

The analysis presented above reminds us of another iron-selenide, layered
KFe$_2$Se$_2$, in which the nominal valence of Fe is $+1.5$ and a two-dimensional block AFM order
exists in each layer \cite{Li:Prb12,*Li:Prl}.
According to the symmetry analysis, each layer of KFe$_2$Se$_2$ should be
FE polarized due to the exchange striction.
However, the FE $P$ cancels between layers,
resulting in an antiferroelectric material.

{\it Prospect.} It is recognized that electron correlations are crucial for high-$T_{\rm C}$ SCs, but they are also equally important in magnetic multiferroics, e.g. to stabilize the $2\times2$ spin block order of BaFe$_2$Se$_3$ \cite{Luo:Prb} that eventually leads to the ferroelectricity discussed here. In fact, the parent materials of high-$T_{\rm C}$ SCs and type-II multiferroics are both antiferromagnets with full or partial Mottness. More in general, the consequences of correlation such as the orbital-selective Mottness \cite{Medici:Prl}, frustrating effects in magnetism, and even strong electron-phonon couplings \cite{Liang:Prl}, all may provide a common fertile environment for both superconductivity and multiferroicity to develop. While it is still an open question to show convincingly whether this leads to cooperation or competition between the two states, BaFe$_2$Se$_3$ establishes a good starting point to explore these ideas.

{\it Summary.} Using a symmetry analysis and first-principles calculations,
the multiferroicity of BaFe$_2$Se$_3$ has been predicted. Different from most
previous magnetic multiferroics, BaFe$_2$Se$_3$ should be ferrielectric but
its net polarization remains large and its critical temperature high.
Its corresponding ferroelectric phase (close in energy) has a
giant polarization. The multiferroic performance of BaFe$_2$Se$_3$ is in the
topmost range in the type-II multiferroic family, making
it an attractive system for further studies. The present
experimental difficulty to obtain a pure phase is here
explained by the covalent bonds scenario. Our study broadens
the research area of multiferroics and leads to a cross fertilization
between superconductors and multiferroics.

We thank K.F. Wang, P. Yu, L. Li, M.F. Liu, J. Neilson, S.-W. Cheong, H.J. Xiang, Q.F. Zhang, Q. Luo, C.L. Zhang, and H. Takahashi for helpful discussions. Work was supported by
the 973 Projects of China (2011CB922101) and NSFC (11274060, 51322206, 11234005). E.D.
was supported by the U.S. DOE, Office of Basic Energy Sciences, Materials
Sciences and Engineering Division.


\begin{thebibliography}{45}%
\makeatletter
\providecommand \@ifxundefined [1]{%
 \@ifx{#1\undefined}
}%
\providecommand \@ifnum [1]{%
 \ifnum #1\expandafter \@firstoftwo
 \else \expandafter \@secondoftwo
 \fi
}%
\providecommand \@ifx [1]{%
 \ifx #1\expandafter \@firstoftwo
 \else \expandafter \@secondoftwo
 \fi
}%
\providecommand \natexlab [1]{#1}%
\providecommand \enquote  [1]{``#1''}%
\providecommand \bibnamefont  [1]{#1}%
\providecommand \bibfnamefont [1]{#1}%
\providecommand \citenamefont [1]{#1}%
\providecommand \href@noop [0]{\@secondoftwo}%
\providecommand \href [0]{\begingroup \@sanitize@url \@href}%
\providecommand \@href[1]{\@@startlink{#1}\@@href}%
\providecommand \@@href[1]{\endgroup#1\@@endlink}%
\providecommand \@sanitize@url [0]{\catcode `\\12\catcode `\$12\catcode
  `\&12\catcode `\#12\catcode `\^12\catcode `\_12\catcode `\%12\relax}%
\providecommand \@@startlink[1]{}%
\providecommand \@@endlink[0]{}%
\providecommand \url  [0]{\begingroup\@sanitize@url \@url }%
\providecommand \@url [1]{\endgroup\@href {#1}{\urlprefix }}%
\providecommand \urlprefix  [0]{URL }%
\providecommand \Eprint [0]{\href }%
\providecommand \doibase [0]{http://dx.doi.org/}%
\providecommand \selectlanguage [0]{\@gobble}%
\providecommand \bibinfo  [0]{\@secondoftwo}%
\providecommand \bibfield  [0]{\@secondoftwo}%
\providecommand \translation [1]{[#1]}%
\providecommand \BibitemOpen [0]{}%
\providecommand \bibitemStop [0]{}%
\providecommand \bibitemNoStop [0]{.\EOS\space}%
\providecommand \EOS [0]{\spacefactor3000\relax}%
\providecommand \BibitemShut  [1]{\csname bibitem#1\endcsname}%
\let\auto@bib@innerbib\@empty
\bibitem [{\citenamefont {Khomskii}(2009)}]{Khomskii:Phy}%
  \BibitemOpen
  \bibfield  {author} {\bibinfo {author} {\bibfnamefont {D.}~\bibnamefont
  {Khomskii}},\ }\href@noop {} {\bibfield  {journal} {\bibinfo  {journal}
  {Physics}\ }\textbf {\bibinfo {volume} {2}},\ \bibinfo {pages} {20} (\bibinfo
  {year} {2009})}\BibitemShut {NoStop}%
\bibitem [{\citenamefont {Cheong}\ and\ \citenamefont
  {Mostovoy}(2007)}]{Cheong:Nm}%
  \BibitemOpen
  \bibfield  {author} {\bibinfo {author} {\bibfnamefont {S.-W.}\ \bibnamefont
  {Cheong}}\ and\ \bibinfo {author} {\bibfnamefont {M.}~\bibnamefont
  {Mostovoy}},\ }\href@noop {} {\bibfield  {journal} {\bibinfo  {journal} {Nat.
  Mater.}\ }\textbf {\bibinfo {volume} {6}},\ \bibinfo {pages} {13} (\bibinfo
  {year} {2007})}\BibitemShut {NoStop}%
\bibitem [{\citenamefont {Wang}\ \emph {et~al.}(2009)\citenamefont {Wang},
  \citenamefont {Liu},\ and\ \citenamefont {Ren}}]{Wang:Ap}%
  \BibitemOpen
  \bibfield  {author} {\bibinfo {author} {\bibfnamefont {K.~F.}\ \bibnamefont
  {Wang}}, \bibinfo {author} {\bibfnamefont {J.-M.}\ \bibnamefont {Liu}}, \
  and\ \bibinfo {author} {\bibfnamefont {Z.~F.}\ \bibnamefont {Ren}},\
  }\href@noop {} {\bibfield  {journal} {\bibinfo  {journal} {Adv. Phys.}\
  }\textbf {\bibinfo {volume} {58}},\ \bibinfo {pages} {321} (\bibinfo {year}
  {2009})}\BibitemShut {NoStop}%
\bibitem [{\citenamefont {Dong}\ and\ \citenamefont {Liu}(2012)}]{Dong:Mplb}%
  \BibitemOpen
  \bibfield  {author} {\bibinfo {author} {\bibfnamefont {S.}~\bibnamefont
  {Dong}}\ and\ \bibinfo {author} {\bibfnamefont {J.-M.}\ \bibnamefont {Liu}},\
  }\href@noop {} {\bibfield  {journal} {\bibinfo  {journal} {Mod. Phys. Lett.
  B}\ }\textbf {\bibinfo {volume} {26}},\ \bibinfo {pages} {1230004} (\bibinfo
  {year} {2012})}\BibitemShut {NoStop}%
\bibitem [{\citenamefont {Zhang}\ \emph {et~al.}(2011)\citenamefont {Zhang},
  \citenamefont {Dong}, \citenamefont {Yan}, \citenamefont {Guo}, \citenamefont
  {Zhang}, \citenamefont {Yunoki}, \citenamefont {Dagotto},\ and\ \citenamefont
  {Liu}}]{Zhang:Prb11}%
  \BibitemOpen
  \bibfield  {author} {\bibinfo {author} {\bibfnamefont {G.~Q.}\ \bibnamefont
  {Zhang}}, \bibinfo {author} {\bibfnamefont {S.}~\bibnamefont {Dong}},
  \bibinfo {author} {\bibfnamefont {Z.~B.}\ \bibnamefont {Yan}}, \bibinfo
  {author} {\bibfnamefont {Y.~Y.}\ \bibnamefont {Guo}}, \bibinfo {author}
  {\bibfnamefont {Q.~F.}\ \bibnamefont {Zhang}}, \bibinfo {author}
  {\bibfnamefont {S.}~\bibnamefont {Yunoki}}, \bibinfo {author} {\bibfnamefont
  {E.}~\bibnamefont {Dagotto}}, \ and\ \bibinfo {author} {\bibfnamefont
  {J.-M.}\ \bibnamefont {Liu}},\ }\href@noop {} {\bibfield  {journal} {\bibinfo
   {journal} {Phys. Rev. B}\ }\textbf {\bibinfo {volume} {84}},\ \bibinfo
  {pages} {174413} (\bibinfo {year} {2011})}\BibitemShut {NoStop}%
\bibitem [{\citenamefont {Johnson}\ \emph {et~al.}(2012)\citenamefont
  {Johnson}, \citenamefont {Chapon}, \citenamefont {Khalyavin}, \citenamefont
  {Manuel}, \citenamefont {Radaelli},\ and\ \citenamefont
  {Martin}}]{Johnson:Prl}%
  \BibitemOpen
  \bibfield  {author} {\bibinfo {author} {\bibfnamefont {R.~D.}\ \bibnamefont
  {Johnson}}, \bibinfo {author} {\bibfnamefont {L.~C.}\ \bibnamefont {Chapon}},
  \bibinfo {author} {\bibfnamefont {D.~D.}\ \bibnamefont {Khalyavin}}, \bibinfo
  {author} {\bibfnamefont {P.}~\bibnamefont {Manuel}}, \bibinfo {author}
  {\bibfnamefont {P.~G.}\ \bibnamefont {Radaelli}}, \ and\ \bibinfo {author}
  {\bibfnamefont {C.}~\bibnamefont {Martin}},\ }\href@noop {} {\bibfield
  {journal} {\bibinfo  {journal} {Phys. Rev. Lett.}\ }\textbf {\bibinfo
  {volume} {108}},\ \bibinfo {pages} {067201} (\bibinfo {year}
  {2012})}\BibitemShut {NoStop}%
\bibitem [{\citenamefont {Lu}\ \emph {et~al.}(2012)\citenamefont {Lu},
  \citenamefont {Whangbo}, \citenamefont {Dong}, \citenamefont {Gong},\ and\
  \citenamefont {Xiang}}]{Lu:Prl}%
  \BibitemOpen
  \bibfield  {author} {\bibinfo {author} {\bibfnamefont {X.~Z.}\ \bibnamefont
  {Lu}}, \bibinfo {author} {\bibfnamefont {M.-H.}\ \bibnamefont {Whangbo}},
  \bibinfo {author} {\bibfnamefont {S.}~\bibnamefont {Dong}}, \bibinfo {author}
  {\bibfnamefont {X.~G.}\ \bibnamefont {Gong}}, \ and\ \bibinfo {author}
  {\bibfnamefont {H.~J.}\ \bibnamefont {Xiang}},\ }\href@noop {} {\bibfield
  {journal} {\bibinfo  {journal} {Phys. Rev. Lett.}\ }\textbf {\bibinfo
  {volume} {108}},\ \bibinfo {pages} {187204} (\bibinfo {year}
  {2012})}\BibitemShut {NoStop}%
\bibitem [{\citenamefont {Dong}\ \emph {et~al.}(2009)\citenamefont {Dong},
  \citenamefont {Yu}, \citenamefont {Liu},\ and\ \citenamefont
  {Dagotto}}]{Dong:Prl}%
  \BibitemOpen
  \bibfield  {author} {\bibinfo {author} {\bibfnamefont {S.}~\bibnamefont
  {Dong}}, \bibinfo {author} {\bibfnamefont {R.}~\bibnamefont {Yu}}, \bibinfo
  {author} {\bibfnamefont {J.-M.}\ \bibnamefont {Liu}}, \ and\ \bibinfo
  {author} {\bibfnamefont {E.}~\bibnamefont {Dagotto}},\ }\href@noop {}
  {\bibfield  {journal} {\bibinfo  {journal} {Phys. Rev. Lett.}\ }\textbf
  {\bibinfo {volume} {103}},\ \bibinfo {pages} {107204} (\bibinfo {year}
  {2009})}\BibitemShut {NoStop}%
\bibitem [{\citenamefont {Perks}\ \emph {et~al.}(2012)\citenamefont {Perks},
  \citenamefont {Johnson}, \citenamefont {Martin}, \citenamefont {Chapon},\
  and\ \citenamefont {Radaelli}}]{Perks:Nc}%
  \BibitemOpen
  \bibfield  {author} {\bibinfo {author} {\bibfnamefont {N.~J.}\ \bibnamefont
  {Perks}}, \bibinfo {author} {\bibfnamefont {R.~D.}\ \bibnamefont {Johnson}},
  \bibinfo {author} {\bibfnamefont {C.}~\bibnamefont {Martin}}, \bibinfo
  {author} {\bibfnamefont {L.~C.}\ \bibnamefont {Chapon}}, \ and\ \bibinfo
  {author} {\bibfnamefont {P.~G.}\ \bibnamefont {Radaelli}},\ }\href@noop {}
  {\bibfield  {journal} {\bibinfo  {journal} {Nat. Commun.}\ }\textbf {\bibinfo
  {volume} {3}},\ \bibinfo {pages} {1277} (\bibinfo {year} {2012})}\BibitemShut
  {NoStop}%
\bibitem [{\citenamefont {Mostovoy}(2012)}]{Mostovoy:Phy}%
  \BibitemOpen
  \bibfield  {author} {\bibinfo {author} {\bibfnamefont {M.}~\bibnamefont
  {Mostovoy}},\ }\href@noop {} {\bibfield  {journal} {\bibinfo  {journal}
  {Physics}\ }\textbf {\bibinfo {volume} {5}},\ \bibinfo {pages} {16} (\bibinfo
  {year} {2012})}\BibitemShut {NoStop}%
\bibitem [{\citenamefont {Kimura}\ \emph {et~al.}(2008)\citenamefont {Kimura},
  \citenamefont {Sekio}, \citenamefont {Nakamura}, \citenamefont {Siegrist},\
  and\ \citenamefont {Ramirez}}]{Kimura:Nm}%
  \BibitemOpen
  \bibfield  {author} {\bibinfo {author} {\bibfnamefont {T.}~\bibnamefont
  {Kimura}}, \bibinfo {author} {\bibfnamefont {Y.}~\bibnamefont {Sekio}},
  \bibinfo {author} {\bibfnamefont {H.}~\bibnamefont {Nakamura}}, \bibinfo
  {author} {\bibfnamefont {T.}~\bibnamefont {Siegrist}}, \ and\ \bibinfo
  {author} {\bibfnamefont {A.~P.}\ \bibnamefont {Ramirez}},\ }\href@noop {}
  {\bibfield  {journal} {\bibinfo  {journal} {Nat. Mater.}\ }\textbf {\bibinfo
  {volume} {7}},\ \bibinfo {pages} {291} (\bibinfo {year} {2008})}\BibitemShut
  {NoStop}%
\bibitem [{\citenamefont {Johnston}(2010)}]{Johnston:Ap}%
  \BibitemOpen
  \bibfield  {author} {\bibinfo {author} {\bibfnamefont {D.~C.}\ \bibnamefont
  {Johnston}},\ }\href@noop {} {\bibfield  {journal} {\bibinfo  {journal} {Adv.
  Phys.}\ }\textbf {\bibinfo {volume} {59}},\ \bibinfo {pages} {803} (\bibinfo
  {year} {2010})}\BibitemShut {NoStop}%
\bibitem [{\citenamefont {Stewart}(2011)}]{Stewart:Rmp}%
  \BibitemOpen
  \bibfield  {author} {\bibinfo {author} {\bibfnamefont {G.~R.}\ \bibnamefont
  {Stewart}},\ }\href@noop {} {\bibfield  {journal} {\bibinfo  {journal} {Rev.
  Mod. Phys.}\ }\textbf {\bibinfo {volume} {83}},\ \bibinfo {pages} {1589}
  (\bibinfo {year} {2011})}\BibitemShut {NoStop}%
\bibitem [{\citenamefont {Dagotto}(2013)}]{Dagotto:Rmp}%
  \BibitemOpen
  \bibfield  {author} {\bibinfo {author} {\bibfnamefont {E.}~\bibnamefont
  {Dagotto}},\ }\href@noop {} {\bibfield  {journal} {\bibinfo  {journal} {Rev.
  Mod. Phys.}\ }\textbf {\bibinfo {volume} {85}},\ \bibinfo {pages} {849}
  (\bibinfo {year} {2013})}\BibitemShut {NoStop}%
\bibitem [{\citenamefont {Krzton-{M}aziopa}\ \emph {et~al.}(2011)\citenamefont
  {Krzton-{M}aziopa}, \citenamefont {Pomjakushina}, \citenamefont
  {Pomjakushin}, \citenamefont {Sheptyakov}, \citenamefont {Chernyshov},
  \citenamefont {Svitlyk},\ and\ \citenamefont {Conder}}]{Maziopa:Jpcm}%
  \BibitemOpen
  \bibfield  {author} {\bibinfo {author} {\bibfnamefont {A.}~\bibnamefont
  {Krzton-{M}aziopa}}, \bibinfo {author} {\bibfnamefont {E.}~\bibnamefont
  {Pomjakushina}}, \bibinfo {author} {\bibfnamefont {V.}~\bibnamefont
  {Pomjakushin}}, \bibinfo {author} {\bibfnamefont {D.}~\bibnamefont
  {Sheptyakov}}, \bibinfo {author} {\bibfnamefont {D.}~\bibnamefont
  {Chernyshov}}, \bibinfo {author} {\bibfnamefont {V.}~\bibnamefont {Svitlyk}},
  \ and\ \bibinfo {author} {\bibfnamefont {K.}~\bibnamefont {Conder}},\
  }\href@noop {} {\bibfield  {journal} {\bibinfo  {journal} {J. Phys.: Condens.
  Matter}\ }\textbf {\bibinfo {volume} {23}},\ \bibinfo {pages} {402201}
  (\bibinfo {year} {2011})}\BibitemShut {NoStop}%
\bibitem [{\citenamefont {Svitlyk}\ \emph {et~al.}(2013)\citenamefont
  {Svitlyk}, \citenamefont {Chernyshov}, \citenamefont {Pomjakushina},
  \citenamefont {Krzton-Maziopa}, \citenamefont {Conder}, \citenamefont
  {Pomjakushin}, \citenamefont {P\"ottgen},\ and\ \citenamefont
  {Dmitriev}}]{Svitlyk:Jpcm}%
  \BibitemOpen
  \bibfield  {author} {\bibinfo {author} {\bibfnamefont {V.}~\bibnamefont
  {Svitlyk}}, \bibinfo {author} {\bibfnamefont {D.}~\bibnamefont {Chernyshov}},
  \bibinfo {author} {\bibfnamefont {E.}~\bibnamefont {Pomjakushina}}, \bibinfo
  {author} {\bibfnamefont {A.}~\bibnamefont {Krzton-Maziopa}}, \bibinfo
  {author} {\bibfnamefont {K.}~\bibnamefont {Conder}}, \bibinfo {author}
  {\bibfnamefont {V.}~\bibnamefont {Pomjakushin}}, \bibinfo {author}
  {\bibfnamefont {R.}~\bibnamefont {P\"ottgen}}, \ and\ \bibinfo {author}
  {\bibfnamefont {V.}~\bibnamefont {Dmitriev}},\ }\href@noop {} {\bibfield
  {journal} {\bibinfo  {journal} {J. Phys.: Condens. Matter}\ }\textbf
  {\bibinfo {volume} {25}},\ \bibinfo {pages} {315403} (\bibinfo {year}
  {2013})}\BibitemShut {NoStop}%
\bibitem [{\citenamefont {Caron}\ \emph {et~al.}(2011)\citenamefont {Caron},
  \citenamefont {Neilson}, \citenamefont {Miller}, \citenamefont {Llobet},\
  and\ \citenamefont {Mc{Q}ueen}}]{Caron:Prb}%
  \BibitemOpen
  \bibfield  {author} {\bibinfo {author} {\bibfnamefont {J.~M.}\ \bibnamefont
  {Caron}}, \bibinfo {author} {\bibfnamefont {J.~R.}\ \bibnamefont {Neilson}},
  \bibinfo {author} {\bibfnamefont {D.~C.}\ \bibnamefont {Miller}}, \bibinfo
  {author} {\bibfnamefont {A.}~\bibnamefont {Llobet}}, \ and\ \bibinfo {author}
  {\bibfnamefont {T.~M.}\ \bibnamefont {Mc{Q}ueen}},\ }\href@noop {} {\bibfield
   {journal} {\bibinfo  {journal} {Phys. Rev. B}\ }\textbf {\bibinfo {volume}
  {84}},\ \bibinfo {pages} {180409(R)} (\bibinfo {year} {2011})}\BibitemShut
  {NoStop}%
\bibitem [{\citenamefont {Caron}\ \emph {et~al.}(2012)\citenamefont {Caron},
  \citenamefont {Neilson}, \citenamefont {Miller}, \citenamefont {Arpino},
  \citenamefont {Llobet},\ and\ \citenamefont {Mc{Q}ueen}}]{Caron:Prb12}%
  \BibitemOpen
  \bibfield  {author} {\bibinfo {author} {\bibfnamefont {J.~M.}\ \bibnamefont
  {Caron}}, \bibinfo {author} {\bibfnamefont {J.~R.}\ \bibnamefont {Neilson}},
  \bibinfo {author} {\bibfnamefont {D.~C.}\ \bibnamefont {Miller}}, \bibinfo
  {author} {\bibfnamefont {K.}~\bibnamefont {Arpino}}, \bibinfo {author}
  {\bibfnamefont {A.}~\bibnamefont {Llobet}}, \ and\ \bibinfo {author}
  {\bibfnamefont {T.~M.}\ \bibnamefont {Mc{Q}ueen}},\ }\href@noop {} {\bibfield
   {journal} {\bibinfo  {journal} {Phys. Rev. B}\ }\textbf {\bibinfo {volume}
  {85}},\ \bibinfo {pages} {180405(R)} (\bibinfo {year} {2012})}\BibitemShut
  {NoStop}%
\bibitem [{\citenamefont {Nambu}\ \emph {et~al.}(2012)\citenamefont {Nambu},
  \citenamefont {Ohgushi}, \citenamefont {Suzuki}, \citenamefont {Du},
  \citenamefont {Avdeev}, \citenamefont {Uwatoko}, \citenamefont {Munakata},
  \citenamefont {Fukazawa}, \citenamefont {Chi}, \citenamefont {Ueda},\ and\
  \citenamefont {Sato}}]{Nambu:Prb}%
  \BibitemOpen
  \bibfield  {author} {\bibinfo {author} {\bibfnamefont {Y.}~\bibnamefont
  {Nambu}}, \bibinfo {author} {\bibfnamefont {K.}~\bibnamefont {Ohgushi}},
  \bibinfo {author} {\bibfnamefont {S.}~\bibnamefont {Suzuki}}, \bibinfo
  {author} {\bibfnamefont {F.}~\bibnamefont {Du}}, \bibinfo {author}
  {\bibfnamefont {M.}~\bibnamefont {Avdeev}}, \bibinfo {author} {\bibfnamefont
  {Y.}~\bibnamefont {Uwatoko}}, \bibinfo {author} {\bibfnamefont
  {K.}~\bibnamefont {Munakata}}, \bibinfo {author} {\bibfnamefont
  {H.}~\bibnamefont {Fukazawa}}, \bibinfo {author} {\bibfnamefont
  {S.}~\bibnamefont {Chi}}, \bibinfo {author} {\bibfnamefont {Y.}~\bibnamefont
  {Ueda}}, \ and\ \bibinfo {author} {\bibfnamefont {T.~J.}\ \bibnamefont
  {Sato}},\ }\href@noop {} {\bibfield  {journal} {\bibinfo  {journal} {Phys.
  Rev. B}\ }\textbf {\bibinfo {volume} {85}},\ \bibinfo {pages} {064413}
  (\bibinfo {year} {2012})}\BibitemShut {NoStop}%
\bibitem [{\citenamefont {Saparov}\ \emph {et~al.}(2011)\citenamefont
  {Saparov}, \citenamefont {Calder}, \citenamefont {Sipos}, \citenamefont
  {Cao}, \citenamefont {Chi}, \citenamefont {Singh}, \citenamefont
  {Christianson}, \citenamefont {Lumsden},\ and\ \citenamefont
  {Sefat}}]{Saparov:Prb}%
  \BibitemOpen
  \bibfield  {author} {\bibinfo {author} {\bibfnamefont {B.}~\bibnamefont
  {Saparov}}, \bibinfo {author} {\bibfnamefont {S.}~\bibnamefont {Calder}},
  \bibinfo {author} {\bibfnamefont {B.}~\bibnamefont {Sipos}}, \bibinfo
  {author} {\bibfnamefont {H.}~\bibnamefont {Cao}}, \bibinfo {author}
  {\bibfnamefont {S.}~\bibnamefont {Chi}}, \bibinfo {author} {\bibfnamefont
  {D.~J.}\ \bibnamefont {Singh}}, \bibinfo {author} {\bibfnamefont {A.~D.}\
  \bibnamefont {Christianson}}, \bibinfo {author} {\bibfnamefont {M.~D.}\
  \bibnamefont {Lumsden}}, \ and\ \bibinfo {author} {\bibfnamefont {A.~S.}\
  \bibnamefont {Sefat}},\ }\href@noop {} {\bibfield  {journal} {\bibinfo
  {journal} {Phys. Rev. B}\ }\textbf {\bibinfo {volume} {84}},\ \bibinfo
  {pages} {245132} (\bibinfo {year} {2011})}\BibitemShut {NoStop}%
\bibitem [{\citenamefont {Medvedev}\ \emph {et~al.}(2012)\citenamefont
  {Medvedev}, \citenamefont {Nekrasov},\ and\ \citenamefont
  {Sadovskii}}]{Medvedev:Jetp}%
  \BibitemOpen
  \bibfield  {author} {\bibinfo {author} {\bibfnamefont {M.~V.}\ \bibnamefont
  {Medvedev}}, \bibinfo {author} {\bibfnamefont {I.~A.}\ \bibnamefont
  {Nekrasov}}, \ and\ \bibinfo {author} {\bibfnamefont {M.~V.}\ \bibnamefont
  {Sadovskii}},\ }\href@noop {} {\bibfield  {journal} {\bibinfo  {journal}
  {JETP Lett.}\ }\textbf {\bibinfo {volume} {95}},\ \bibinfo {pages} {37}
  (\bibinfo {year} {2012})}\BibitemShut {NoStop}%
\bibitem [{\citenamefont {Lei}\ \emph {et~al.}(2011)\citenamefont {Lei},
  \citenamefont {Ryu}, \citenamefont {Frenkel},\ and\ \citenamefont
  {Petrovic}}]{Lei:Prb}%
  \BibitemOpen
  \bibfield  {author} {\bibinfo {author} {\bibfnamefont {H.}~\bibnamefont
  {Lei}}, \bibinfo {author} {\bibfnamefont {H.}~\bibnamefont {Ryu}}, \bibinfo
  {author} {\bibfnamefont {A.~I.}\ \bibnamefont {Frenkel}}, \ and\ \bibinfo
  {author} {\bibfnamefont {C.}~\bibnamefont {Petrovic}},\ }\href@noop {}
  {\bibfield  {journal} {\bibinfo  {journal} {Phys. Rev. B}\ }\textbf {\bibinfo
  {volume} {84}},\ \bibinfo {pages} {214511} (\bibinfo {year}
  {2011})}\BibitemShut {NoStop}%
\bibitem [{\citenamefont {Luo}\ \emph {et~al.}(2013)\citenamefont {Luo},
  \citenamefont {Nicholson}, \citenamefont {Rinc\'on}, \citenamefont {Liang},
  \citenamefont {Riera}, \citenamefont {Alvarez}, \citenamefont {Wang},
  \citenamefont {Ku}, \citenamefont {Samolyuk}, \citenamefont {Moreo},\ and\
  \citenamefont {Dagotto}}]{Luo:Prb}%
  \BibitemOpen
  \bibfield  {author} {\bibinfo {author} {\bibfnamefont {Q.}~\bibnamefont
  {Luo}}, \bibinfo {author} {\bibfnamefont {A.}~\bibnamefont {Nicholson}},
  \bibinfo {author} {\bibfnamefont {J.}~\bibnamefont {Rinc\'on}}, \bibinfo
  {author} {\bibfnamefont {S.}~\bibnamefont {Liang}}, \bibinfo {author}
  {\bibfnamefont {J.}~\bibnamefont {Riera}}, \bibinfo {author} {\bibfnamefont
  {G.}~\bibnamefont {Alvarez}}, \bibinfo {author} {\bibfnamefont
  {L.}~\bibnamefont {Wang}}, \bibinfo {author} {\bibfnamefont {W.}~\bibnamefont
  {Ku}}, \bibinfo {author} {\bibfnamefont {G.~D.}\ \bibnamefont {Samolyuk}},
  \bibinfo {author} {\bibfnamefont {A.}~\bibnamefont {Moreo}}, \ and\ \bibinfo
  {author} {\bibfnamefont {E.}~\bibnamefont {Dagotto}},\ }\href@noop {}
  {\bibfield  {journal} {\bibinfo  {journal} {Phys. Rev. B}\ }\textbf {\bibinfo
  {volume} {87}},\ \bibinfo {pages} {024404} (\bibinfo {year}
  {2013})}\BibitemShut {NoStop}%
\bibitem [{\citenamefont {Sergienko}\ \emph {et~al.}(2006)\citenamefont
  {Sergienko}, \citenamefont {\c{S}en},\ and\ \citenamefont
  {Dagotto}}]{Sergienko:Prl}%
  \BibitemOpen
  \bibfield  {author} {\bibinfo {author} {\bibfnamefont {I.~A.}\ \bibnamefont
  {Sergienko}}, \bibinfo {author} {\bibfnamefont {C.}~\bibnamefont {\c{S}en}},
  \ and\ \bibinfo {author} {\bibfnamefont {E.}~\bibnamefont {Dagotto}},\
  }\href@noop {} {\bibfield  {journal} {\bibinfo  {journal} {Phys. Rev. Lett.}\
  }\textbf {\bibinfo {volume} {97}},\ \bibinfo {pages} {227204} (\bibinfo
  {year} {2006})}\BibitemShut {NoStop}%
\bibitem [{\citenamefont {Choi}\ \emph {et~al.}(2008)\citenamefont {Choi},
  \citenamefont {Yi}, \citenamefont {Lee}, \citenamefont {Huang}, \citenamefont
  {Kiryukhin},\ and\ \citenamefont {Cheong}}]{Choi:Prl}%
  \BibitemOpen
  \bibfield  {author} {\bibinfo {author} {\bibfnamefont {Y.~J.}\ \bibnamefont
  {Choi}}, \bibinfo {author} {\bibfnamefont {H.~T.}\ \bibnamefont {Yi}},
  \bibinfo {author} {\bibfnamefont {S.}~\bibnamefont {Lee}}, \bibinfo {author}
  {\bibfnamefont {Q.}~\bibnamefont {Huang}}, \bibinfo {author} {\bibfnamefont
  {V.}~\bibnamefont {Kiryukhin}}, \ and\ \bibinfo {author} {\bibfnamefont
  {S.-W.}\ \bibnamefont {Cheong}},\ }\href@noop {} {\bibfield  {journal}
  {\bibinfo  {journal} {Phys. Rev. Lett.}\ }\textbf {\bibinfo {volume} {100}},\
  \bibinfo {pages} {047601} (\bibinfo {year} {2008})}\BibitemShut {NoStop}%
\bibitem [{\citenamefont {Bousquet}\ \emph {et~al.}(2008)\citenamefont
  {Bousquet}, \citenamefont {Dawber}, \citenamefont {Stucki}, \citenamefont
  {Lichtensteiger}, \citenamefont {Hermet}, \citenamefont {Gariglio},
  \citenamefont {Triscone},\ and\ \citenamefont {Ghosez}}]{Bousquet:Nat}%
  \BibitemOpen
  \bibfield  {author} {\bibinfo {author} {\bibfnamefont {E.}~\bibnamefont
  {Bousquet}}, \bibinfo {author} {\bibfnamefont {M.}~\bibnamefont {Dawber}},
  \bibinfo {author} {\bibfnamefont {N.}~\bibnamefont {Stucki}}, \bibinfo
  {author} {\bibfnamefont {C.}~\bibnamefont {Lichtensteiger}}, \bibinfo
  {author} {\bibfnamefont {P.}~\bibnamefont {Hermet}}, \bibinfo {author}
  {\bibfnamefont {S.}~\bibnamefont {Gariglio}}, \bibinfo {author}
  {\bibfnamefont {J.-M.}\ \bibnamefont {Triscone}}, \ and\ \bibinfo {author}
  {\bibfnamefont {P.}~\bibnamefont {Ghosez}},\ }\href@noop {} {\bibfield
  {journal} {\bibinfo  {journal} {Nature (London)}\ }\textbf {\bibinfo {volume}
  {452}},\ \bibinfo {pages} {732} (\bibinfo {year} {2008})}\BibitemShut
  {NoStop}%
\bibitem [{\citenamefont {Benedek}\ and\ \citenamefont
  {Fennie}(2011)}]{Benedek:Prl}%
  \BibitemOpen
  \bibfield  {author} {\bibinfo {author} {\bibfnamefont {N.~A.}\ \bibnamefont
  {Benedek}}\ and\ \bibinfo {author} {\bibfnamefont {C.~J.}\ \bibnamefont
  {Fennie}},\ }\href@noop {} {\bibfield  {journal} {\bibinfo  {journal} {Phys.
  Rev. Lett.}\ }\textbf {\bibinfo {volume} {106}},\ \bibinfo {pages} {107204}
  (\bibinfo {year} {2011})}\BibitemShut {NoStop}%
\bibitem [{\citenamefont {Yang}\ \emph {et~al.}(2014)\citenamefont {Yang},
  \citenamefont {{\'I}{\~n}iguez}, \citenamefont {Mao},\ and\ \citenamefont
  {Bellaiche}}]{Yang:Prl}%
  \BibitemOpen
  \bibfield  {author} {\bibinfo {author} {\bibfnamefont {Y.}~\bibnamefont
  {Yang}}, \bibinfo {author} {\bibfnamefont {J.}~\bibnamefont
  {{\'I}{\~n}iguez}}, \bibinfo {author} {\bibfnamefont {A.-J.}\ \bibnamefont
  {Mao}}, \ and\ \bibinfo {author} {\bibfnamefont {L.}~\bibnamefont
  {Bellaiche}},\ }\href@noop {} {\bibfield  {journal} {\bibinfo  {journal}
  {Phys. Rev. Lett.}\ }\textbf {\bibinfo {volume} {112}},\ \bibinfo {pages}
  {057202} (\bibinfo {year} {2014})}\BibitemShut {NoStop}%
\bibitem [{Note1()}]{Note1}%
  \BibitemOpen
  \bibinfo {note} {Note that all the ladders, either A or B, have the same
  staggered pattern of Se atoms above and below the ladder planes, i.e. a
  plaquette with a Se(5) above the ladder plane for ladder A, corresponds to
  another Se(5) above the ladder plane for ladder B, for the same
  plaquette.}\BibitemShut {Stop}%
\bibitem [{Note2()}]{Note2}%
  \BibitemOpen
  \bibinfo {note} {Using the experimental structure (space group No. 62,
  $Pnma$, orthorhombic) plus the particular magnetic order, the magnetic space
  group becomes monoclinic: (1) No. 14, $P21/c$ for the Cx phase; (2) No. 9,
  $Cc$ for the Block-EX phase; (3) No. 8, $Cm$ for the Block-MF phase. The
  point group of $P21/c$ is $2/m$ which is a nonpolar point group. Then, $P$ is
  forbidden in this group. The point group of both $Cc$ and $Cm$ is $m$, which
  is a polar point group and allows $P$.}\BibitemShut {Stop}%
\bibitem [{\citenamefont {{See Supplemental Material, which includes Refs.
  [20-24]}}()}]{Supp}%
  \BibitemOpen
  \bibfield  {author} {\bibinfo {author} {\bibnamefont {{See Supplemental
  Material, which includes Refs. [20-24]}}}\ }\href@noop {} {}\BibitemShut
  {NoStop}%
\bibitem [{\citenamefont {Bl\"{o}chl}\ \emph {et~al.}(1994)\citenamefont
  {Bl\"{o}chl}, \citenamefont {Jepsen},\ and\ \citenamefont
  {Andersen}}]{Blochl:Prb}%
  \BibitemOpen
  \bibfield  {author} {\bibinfo {author} {\bibfnamefont {P.~E.}\ \bibnamefont
  {Bl\"{o}chl}}, \bibinfo {author} {\bibfnamefont {O.}~\bibnamefont {Jepsen}},
  \ and\ \bibinfo {author} {\bibfnamefont {O.~K.}\ \bibnamefont {Andersen}},\
  }\href@noop {} {\bibfield  {journal} {\bibinfo  {journal} {Phys. Rev. B}\
  }\textbf {\bibinfo {volume} {49}},\ \bibinfo {pages} {16223} (\bibinfo {year}
  {1994})}\BibitemShut {NoStop}%
\bibitem [{\citenamefont {Kresse}\ and\ \citenamefont
  {Hafner}(1993)}]{Kresse:Prb}%
  \BibitemOpen
  \bibfield  {author} {\bibinfo {author} {\bibfnamefont {G.}~\bibnamefont
  {Kresse}}\ and\ \bibinfo {author} {\bibfnamefont {J.}~\bibnamefont
  {Hafner}},\ }\href@noop {} {\bibfield  {journal} {\bibinfo  {journal} {Phys.
  Rev. B}\ }\textbf {\bibinfo {volume} {47}},\ \bibinfo {pages} {558} (\bibinfo
  {year} {1993})}\BibitemShut {NoStop}%
\bibitem [{\citenamefont {Kresse}\ and\ \citenamefont
  {Furthm\"{u}ller}(1996)}]{Kresse:Prb96}%
  \BibitemOpen
  \bibfield  {author} {\bibinfo {author} {\bibfnamefont {G.}~\bibnamefont
  {Kresse}}\ and\ \bibinfo {author} {\bibfnamefont {J.}~\bibnamefont
  {Furthm\"{u}ller}},\ }\href@noop {} {\bibfield  {journal} {\bibinfo
  {journal} {Phys. Rev. B}\ }\textbf {\bibinfo {volume} {54}},\ \bibinfo
  {pages} {11169} (\bibinfo {year} {1996})}\BibitemShut {NoStop}%
\bibitem [{\citenamefont {Dudarev}\ \emph {et~al.}(1998)\citenamefont
  {Dudarev}, \citenamefont {Botton}, \citenamefont {Savrasov}, \citenamefont
  {Humphreys},\ and\ \citenamefont {Sutton}}]{Dudarev:Prb}%
  \BibitemOpen
  \bibfield  {author} {\bibinfo {author} {\bibfnamefont {S.~L.}\ \bibnamefont
  {Dudarev}}, \bibinfo {author} {\bibfnamefont {G.~A.}\ \bibnamefont {Botton}},
  \bibinfo {author} {\bibfnamefont {S.~Y.}\ \bibnamefont {Savrasov}}, \bibinfo
  {author} {\bibfnamefont {C.~J.}\ \bibnamefont {Humphreys}}, \ and\ \bibinfo
  {author} {\bibfnamefont {A.~P.}\ \bibnamefont {Sutton}},\ }\href@noop {}
  {\bibfield  {journal} {\bibinfo  {journal} {Phys. Rev. B}\ }\textbf {\bibinfo
  {volume} {57}},\ \bibinfo {pages} {1505} (\bibinfo {year}
  {1998})}\BibitemShut {NoStop}%
\bibitem [{\citenamefont {King-Smith}\ and\ \citenamefont
  {Vanderbilt}(1993)}]{Smith:Prb}%
  \BibitemOpen
  \bibfield  {author} {\bibinfo {author} {\bibfnamefont {R.~D.}\ \bibnamefont
  {King-Smith}}\ and\ \bibinfo {author} {\bibfnamefont {D.}~\bibnamefont
  {Vanderbilt}},\ }\href@noop {} {\bibfield  {journal} {\bibinfo  {journal}
  {Phys. Rev. B}\ }\textbf {\bibinfo {volume} {47}},\ \bibinfo {pages} {1651}
  (\bibinfo {year} {1993})}\BibitemShut {NoStop}%
\bibitem [{\citenamefont {Picozzi}\ \emph {et~al.}(2007)\citenamefont
  {Picozzi}, \citenamefont {Yamauchi}, \citenamefont {Sanyal}, \citenamefont
  {Sergienko},\ and\ \citenamefont {Dagotto}}]{Picozzi:Prl}%
  \BibitemOpen
  \bibfield  {author} {\bibinfo {author} {\bibfnamefont {S.}~\bibnamefont
  {Picozzi}}, \bibinfo {author} {\bibfnamefont {K.}~\bibnamefont {Yamauchi}},
  \bibinfo {author} {\bibfnamefont {B.}~\bibnamefont {Sanyal}}, \bibinfo
  {author} {\bibfnamefont {I.~A.}\ \bibnamefont {Sergienko}}, \ and\ \bibinfo
  {author} {\bibfnamefont {E.}~\bibnamefont {Dagotto}},\ }\href@noop {}
  {\bibfield  {journal} {\bibinfo  {journal} {Phys. Rev. Lett.}\ }\textbf
  {\bibinfo {volume} {99}},\ \bibinfo {pages} {227201} (\bibinfo {year}
  {2007})}\BibitemShut {NoStop}%
\bibitem [{\citenamefont {Nakamura}\ \emph {et~al.}(2011)\citenamefont
  {Nakamura}, \citenamefont {Tokunaga}, \citenamefont {Kawasaki},\ and\
  \citenamefont {Tokura}}]{Nakamura:Apl}%
  \BibitemOpen
  \bibfield  {author} {\bibinfo {author} {\bibfnamefont {M.}~\bibnamefont
  {Nakamura}}, \bibinfo {author} {\bibfnamefont {Y.}~\bibnamefont {Tokunaga}},
  \bibinfo {author} {\bibfnamefont {M.}~\bibnamefont {Kawasaki}}, \ and\
  \bibinfo {author} {\bibfnamefont {Y.}~\bibnamefont {Tokura}},\ }\href@noop {}
  {\bibfield  {journal} {\bibinfo  {journal} {Appl. Phys. Lett.}\ }\textbf
  {\bibinfo {volume} {98}},\ \bibinfo {pages} {082902} (\bibinfo {year}
  {2011})}\BibitemShut {NoStop}%
\bibitem [{\citenamefont {Kimura}\ \emph {et~al.}(2005)\citenamefont {Kimura},
  \citenamefont {Lawes}, \citenamefont {Goto}, \citenamefont {Tokura},\ and\
  \citenamefont {Ramirez}}]{Kimura:Prb05}%
  \BibitemOpen
  \bibfield  {author} {\bibinfo {author} {\bibfnamefont {T.}~\bibnamefont
  {Kimura}}, \bibinfo {author} {\bibfnamefont {G.}~\bibnamefont {Lawes}},
  \bibinfo {author} {\bibfnamefont {T.}~\bibnamefont {Goto}}, \bibinfo {author}
  {\bibfnamefont {Y.}~\bibnamefont {Tokura}}, \ and\ \bibinfo {author}
  {\bibfnamefont {A.~P.}\ \bibnamefont {Ramirez}},\ }\href@noop {} {\bibfield
  {journal} {\bibinfo  {journal} {Phys. Rev. B}\ }\textbf {\bibinfo {volume}
  {71}},\ \bibinfo {pages} {224425} (\bibinfo {year} {2005})}\BibitemShut
  {NoStop}%
\bibitem [{\citenamefont {Hong}\ and\ \citenamefont
  {Steinfink}(1972)}]{Hong:Jssc}%
  \BibitemOpen
  \bibfield  {author} {\bibinfo {author} {\bibfnamefont {H.~Y.}\ \bibnamefont
  {Hong}}\ and\ \bibinfo {author} {\bibfnamefont {H.}~\bibnamefont
  {Steinfink}},\ }\href@noop {} {\bibfield  {journal} {\bibinfo  {journal} {J.
  Solid State Chem.}\ }\textbf {\bibinfo {volume} {5}},\ \bibinfo {pages} {93}
  (\bibinfo {year} {1972})}\BibitemShut {NoStop}%
\bibitem [{Note3()}]{Note3}%
  \BibitemOpen
  \bibinfo {note} {H. Takahashi, private communication.}\BibitemShut {Stop}%
\bibitem [{\citenamefont {Li}\ \emph {et~al.}(2012{\natexlab{a}})\citenamefont
  {Li}, \citenamefont {Dong}, \citenamefont {Fang},\ and\ \citenamefont
  {Hu}}]{Li:Prb12}%
  \BibitemOpen
  \bibfield  {author} {\bibinfo {author} {\bibfnamefont {W.}~\bibnamefont
  {Li}}, \bibinfo {author} {\bibfnamefont {S.}~\bibnamefont {Dong}}, \bibinfo
  {author} {\bibfnamefont {C.}~\bibnamefont {Fang}}, \ and\ \bibinfo {author}
  {\bibfnamefont {J.~P.}\ \bibnamefont {Hu}},\ }\href@noop {} {\bibfield
  {journal} {\bibinfo  {journal} {Phys. Rev. B}\ }\textbf {\bibinfo {volume}
  {85}},\ \bibinfo {pages} {100407(R)} (\bibinfo {year}
  {2012}{\natexlab{a}})}\BibitemShut {NoStop}%
\bibitem [{\citenamefont {Li}\ \emph {et~al.}(2012{\natexlab{b}})\citenamefont
  {Li}, \citenamefont {Ding}, \citenamefont {Li}, \citenamefont {Deng},
  \citenamefont {Chang}, \citenamefont {He}, \citenamefont {Ji}, \citenamefont
  {Wang}, \citenamefont {Ma}, \citenamefont {Hu}, \citenamefont {Chen},\ and\
  \citenamefont {Xue}}]{Li:Prl}%
  \BibitemOpen
  \bibfield  {author} {\bibinfo {author} {\bibfnamefont {W.}~\bibnamefont
  {Li}}, \bibinfo {author} {\bibfnamefont {H.}~\bibnamefont {Ding}}, \bibinfo
  {author} {\bibfnamefont {Z.}~\bibnamefont {Li}}, \bibinfo {author}
  {\bibfnamefont {P.}~\bibnamefont {Deng}}, \bibinfo {author} {\bibfnamefont
  {K.}~\bibnamefont {Chang}}, \bibinfo {author} {\bibfnamefont
  {K.}~\bibnamefont {He}}, \bibinfo {author} {\bibfnamefont {S.}~\bibnamefont
  {Ji}}, \bibinfo {author} {\bibfnamefont {L.}~\bibnamefont {Wang}}, \bibinfo
  {author} {\bibfnamefont {X.}~\bibnamefont {Ma}}, \bibinfo {author}
  {\bibfnamefont {J.-P.}\ \bibnamefont {Hu}}, \bibinfo {author} {\bibfnamefont
  {X.}~\bibnamefont {Chen}}, \ and\ \bibinfo {author} {\bibfnamefont {Q.-K.}\
  \bibnamefont {Xue}},\ }\href@noop {} {\bibfield  {journal} {\bibinfo
  {journal} {Phys. Rev. Lett.}\ }\textbf {\bibinfo {volume} {109}},\ \bibinfo
  {pages} {057003} (\bibinfo {year} {2012}{\natexlab{b}})}\BibitemShut
  {NoStop}%
\bibitem [{\citenamefont {{de$'$ Medici}}\ \emph {et~al.}(2014)\citenamefont
  {{de$'$ Medici}}, \citenamefont {Giovannetti},\ and\ \citenamefont
  {Capone}}]{Medici:Prl}%
  \BibitemOpen
  \bibfield  {author} {\bibinfo {author} {\bibfnamefont {L.}~\bibnamefont
  {{de$'$ Medici}}}, \bibinfo {author} {\bibfnamefont {G.}~\bibnamefont
  {Giovannetti}}, \ and\ \bibinfo {author} {\bibfnamefont {M.}~\bibnamefont
  {Capone}},\ }\href@noop {} {\bibfield  {journal} {\bibinfo  {journal} {Phys.
  Rev. Lett.}\ }\textbf {\bibinfo {volume} {112}},\ \bibinfo {pages} {177001}
  (\bibinfo {year} {2014})}\BibitemShut {NoStop}%
\bibitem [{\citenamefont {Liang}\ \emph {et~al.}(2013)\citenamefont {Liang},
  \citenamefont {Moreo},\ and\ \citenamefont {Dagotto}}]{Liang:Prl}%
  \BibitemOpen
  \bibfield  {author} {\bibinfo {author} {\bibfnamefont {S.}~\bibnamefont
  {Liang}}, \bibinfo {author} {\bibfnamefont {A.}~\bibnamefont {Moreo}}, \ and\
  \bibinfo {author} {\bibfnamefont {E.}~\bibnamefont {Dagotto}},\ }\href@noop
  {} {\bibfield  {journal} {\bibinfo  {journal} {Phys. Rev. Lett.}\ }\textbf
  {\bibinfo {volume} {111}},\ \bibinfo {pages} {047004} (\bibinfo {year}
  {2013})}\BibitemShut {NoStop}%
\end{thebibliography}
\end{document}